# MITAO: a tool for enabling scholars in the Humanities to use Topic Modelling in their studies

Ivan Heibi[1], Silvio Peroni[2], Luca Pareschi[3], Paolo Ferri[4]

[1] Research Centre for Open Scholarly Metadata, Department of Classical Philology and Italian Studies, University of Bologna, Bologna, Italy
Digital Humanities Advanced Research Centre, Department of Classical Philology and Italian Studies, University of Bologna, Bologna, Italy – ivan.heibi2@unibo.it

[2] Research Centre for Open Scholarly Metadata, Department of Classical Philology and Italian Studies, University of Bologna, Bologna, Italy
Digital Humanities Advanced Research Centre, Department of Classical Philology and Italian Studies, University of Bologna, Bologna, Italy – silvio.peroni@unibo.it

[3] Department of Management and Law, University of Rome Tor Vergata, Rome, Italy – luca.pareschi@uniroma2.it

[4] Department of Management, University of Bologna, Bologna, Italy – p.ferri@unibo.it

**ABSTRACT**
Automatic text analysis methods, such as Topic Modelling, are gaining much attention in Humanities. However, scholars need to have extensive coding skills to use such methods appropriately. The need of having this technical expertise prevents the broad adoption of these methods in Humanities research. In this paper, to help scholars in the Humanities to use Topic Modelling having no or limited coding skills, we introduce MITAO, a web-based tool that allow the definition of a visual workflow which embeds various automatic text analysis operations and allows one to store and share both the workflow and the results of its execution to other researchers, which enables the reproducibility of the analysis. We present an example of an application of use of Topic Modelling with MITAO using a collection of English abstracts of the articles published in "Umanistica Digitale". The results returned by MITAO are shown with dynamic web-based visualizations, which allowed us to have preliminary insights about the evolution of the topics treated over the time in the articles published in "Umanistica Digitale". All the results along with the defined workflows are published and accessible for further studies.

## 1. INTRODUCTION

The quantitative analysis of textual contents is an approach that has been adopted since the end of the 19th century, when John Gilmer Speed, a former New York World editor, used quantitative methods to compare the content of four New York dailies published between 1881 and 1893 [15]. These methods (and, more generally, the recent computational text mining techniques) are considered nowadays an interdisciplinary topic with relations with computer science, statistics, linguistics, sociology, and other social sciences. Today we have massive digital archives (e.g. Project Gutenberg and Google Books) and new computational tools and sophisticated systems for studying them.

In the past years, such computational tools have attracted scholars from the Humanities and Social Sciences domains. Indeed, an increasing number of scholars in the Humanities have started to take advantage of potentials of such computational tools for text processing and analysis [10]. In particular, within the Digital Humanities, such tools are crucial in addressing research questions [8], as also demonstrated by the recently-born Computational Humanities Research community (https://www.computational-humanities-research.org/).

Topic Modelling is one of the textual processing methodologies mentioned above. Topic Modelling aims at determining what events or concepts a document or a collection of documents is discussing [16], something which may have several applications in the Humanities [2] and the Digital Humanities [11].

Technically speaking, Topic Modelling is an automatic technique based on Bayesian statistics, and in particular on an algorithm called Latent Dirichlet Allocation [9]. Topic Modeling analyses texts and creates 'topics', which are bags of words that often co-occur together in the original texts [12]. The underlying idea is that the algorithm is able to elicit a latent structure of topics that constitute the texts in a corpus [9]. As a result, all the words in the textual sources are coded to a topic, and all the original texts are constituted by several topics according to different percentages. Topics can be ontologically different [4], as they can be interpreted as themes, or discourse, or frames [5]. The work done by DiMaggio, Nag, and Blei [3] identifies four relevant characteristics of the Topic modelling analysis that helps researchers cover the limitations arisen from their previous approaches (mainly based on a manual approach): (a) the generated data are available for other researchers willing to test the interpretations and conclusions inferred; (b) the automatic analysis permits dealing with large amounts of documents; (c) the process is an inductive one, such that it allows researchers to discover the structure of the corpus before imposing their priors on the analysis, and also enables the usage of the same corpus to pursue other different research questions; (d) it deals with polysemy, such that it is able to recognize that the meaning of a term depends on its context (therefore the same term might have different meanings depending on its context).

While useful as an approach per se, one has to master at least a programming language and a tool for a visual representation of data to use Topic Modelling correctly. However, scholars in the Humanities are not usually trained in these skills – and, thus, the adoption of Topic Modelling in their research is perceived as an insuperable barrier. Interfaces and visual tools are needed in order to mediate appropriately the use of Topic Modelling, so as to hide its technical implementation to non-experts.

In this paper, we introduce MITAO, a tool which makes Topic Modelling techniques reusable by scholars with no or limited coding skills by means of a visual interface which enables them to create a workflow for processing textual content [6]. MITAO uses the Latent Dirichlet Allocation (LDA) Topic Modelling, one of the earliest and well-known methods, originally developed by Blei, Ng, and Jordan [1]. Any workflow produced using MITAO can be stored and shared among scholars for reproducibility purposes.

In Section 2, we provide an introduction to MITAO, we give an overview of its architecture, we discuss its main features available, and we introduce how to perform a Topic Modelling analysis and to get the tabular and graphical results using it. In Section 3, we show an example of application of MITAO using a collection of the abstracts of the articles published in "Umanistica Digitale" (https://umanisticadigitale.unibo.it/). In Section 4, we conclude this article sketching out some future works.

## 2. MITAO

MITAO (Mashup Interface for Text Analysis Operations) is an open source, user-friendly, modular, and flexible software written in Python and Javascript for performing several kinds of text analysis. MITAO can be run locally on a machine by using any modern Web browser. MITAO is licensed under the ISC License and source code and documentation are available on GitHub at https://github.com/catarsi/mitao.

We developed MITAO to help scholars with no or limited skills in coding to overcome two of the main issues they have to deal with: (a) using computational text analysis techniques for their own research using a particular programming language, and (b) describing and discussing the technical aspects of their analysis. To overcome these limitations, the current version of MITAO (downloadable from its GitHub repository) can:

- convert documents (from PDF to TXT);
- clean textual content (e.g. stopword removal or removal of parts of text through the use of regular expressions);
- perform Topic Modelling;
- provide a quantitative measure of the results (through perplexity score and topic coherence [13]);
- visualize the topic model created with dynamic web-based visualizations;
- save the data and visualizations produced.

In addition, users can save the workflow defined in MITAO and afterwards share it with other colleagues or publish it, so as to foster a reproducibility of the results of a research.

The GUI of MITAO is simple and user friendly. The defined workflow is represented as a graph network composed by two types of nodes: "tool" and "data". A "tool" node implements operations one can run on data, such as a filter (e.g. filtering a document from text values that follow a specific regular expression), a text analysis (e.g. corpus tokenizer), or a terminal operator (e.g. charts and web-based visualizations). Instead, a "data" node represents a single textual file or a collection of textual files (in plain text, PDF, or textual tabular format).

In this paper, we focus on the features of MITAO strictly related to the Topic Modelling analysis.

## 2.1 BUILDING A TOPIC MODELLING

A standard Topic Modelling workflow can be defined according to three main steps: (a) tokenization, (b) building the corpus and dictionary, and (c) building the topic model. In MITAO, these three steps can be defined as shown in Figure 1. The workflow starts with the two "data" nodes which represents a collection of documents (i.e. *docs*) and a list of stopwords (i.e. *stopwords*). Both such nodes are specified as input to the "tool" node *tokenizer* which converts the texts into a list of terms with no stopwords in it. Then, the workflow creates the corpus and dictionary (i.e. "tool" node *corpus and dict builder*) to be further used as input in the creation of the topic model (i.e. "tool" node *lda topic modelling*).

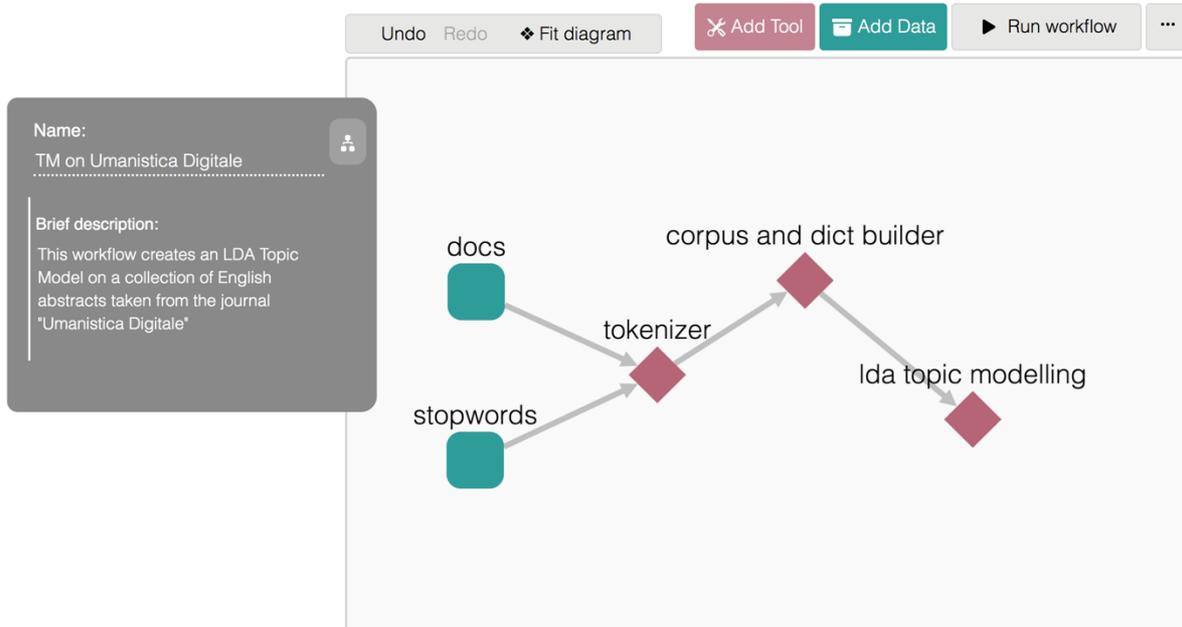

**Figure 1. A Topic Modelling workflow defined in MITAO. Starting from a collection of documents (i.e. *docs*) and a list of stopwords (i.e. *stopwords*), the workflow goes through three different steps: (a) tokenization (i.e. *tokenizer*), (b) building the corpus and dictionary (i.e. *corpus and dict builder*), and finally (c) the creation of the topic model (i.e. *lda topic modelling*).**

## 2.2 TOPIC MODELLING RESULTS

Using MITAO we can generate two important tabular datasets: (a) *termsXtopics*, i.e. the 30 terms that better characterise each topic, and (b) *docsXtopics*, i.e. a list of all the documents of the corpus with their corresponding representativeness for each topic in the topic model we built. Along with the tabular datasets, we can use MITAO to generate two other web-based dynamic visualizations: *LDAvis* and *MTMvis*. LDAvis [14] provides a graphical overview of the topics of our topic model. Such topics are shown in a two-dimensional plane whose centers are determined by computing the distance between topics. MTMvis has been built for MITAO and shows the topic representativeness in the document corpus based on a metadata attribute of such documents. These visualizations enable us to visually investigate the document corpus. In the next section we present a real application of MITAO and demonstrate the potentials of these visualizations.

## 3. AN APPLICATION

In this section we show an example of a Topic Modelling analysis performed using MITAO, accompanied by the datasets and visualizations we obtained by running the workflow. The topic modelling is done on a collection of 51 abstracts in English of the articles published in "Umanistica Digitale". The MITAO workflow we developed, the datasets and the visualizations we obtained are available in [7].

We choose to create a topic model with five different topics. The number of topics should be given as input of the *lda topic modelling* step, along with the dictionary and the corpus expressed as a bag of vectors, created in the *corpus and dict builder* step. The process of choosing this number of topics is out-of-scope of the present paper and have been calculated with the help of a tool MITAO makes available to compute the coherence score of several topic models.

In Figure 2, we show the LDAvis generated as a result of Topic Modelling activity. The chart shows five circles (topics) and, when selecting one of the topics, it shows the 30 most recurrent terms of such topic on the right side of the visualization. Figure 3 shows MTMvis, which plots the distribution of the five topics in time, considering the year of publication of the articles in our corpus.

The combination of both these visualizations can let us come up with some initial insights. From the MTMvis, we see that topic-1 (in blue color) appeared only in 2019. Moving the cursor over such a slice, MTMvis shows that 15.63% of the documents published in 2019 had this as the dominant topic. If we check topic-1 on LDAvis, we see that it has many terms related to the second World War and the Holocaust, such as: "jew", "holocaust", "social", "testimoni", "oorlogsbronnen", "state" etc. Through LDAvis, we can clearly see an intersection between topic-4 and topic-2. While looking at the most recurrent words of these two topics, we notice that both mention words close to the Italian literature. More precisely we can see that topic-4 has a strong relation with the Italian poetry, especially with Dante Alighieri, with words such as "alighieri", "literatur", "librari", "philolog", "poetri", "poet", etc. MTMvis shows that topic-4 (in brown color) had a strong relevance in 2017, and gained less relevance in the following years. Another emerging fact is that topic-5 (in orange color) had a constant relevancy throughout the years (average value of 25%). From LDAvis, we can observe that topic-5 contained words such as "visual", "document", "model", "corpora", "ontolog", etc. From these bags of words, we might infer that topic-5 is related to works dealing with data analysis involving in some cases the definition of a model or dealing with corpora. The constant relevancy of topic-5 along the years make us believe that these subjects are a regular and important part of "Umanistica Digitale" publications.

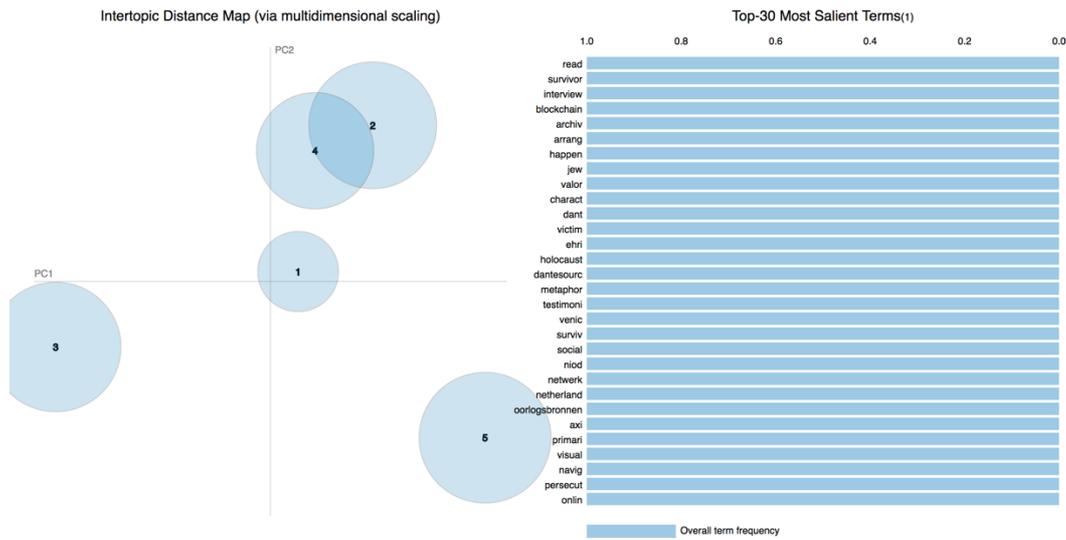

**Figure 2. The LDAvis visualization of the topic model built over a collection of abstracts of articles published in "Umanistica Digitale". This view has been generated using MITAO and it's available in [7].**

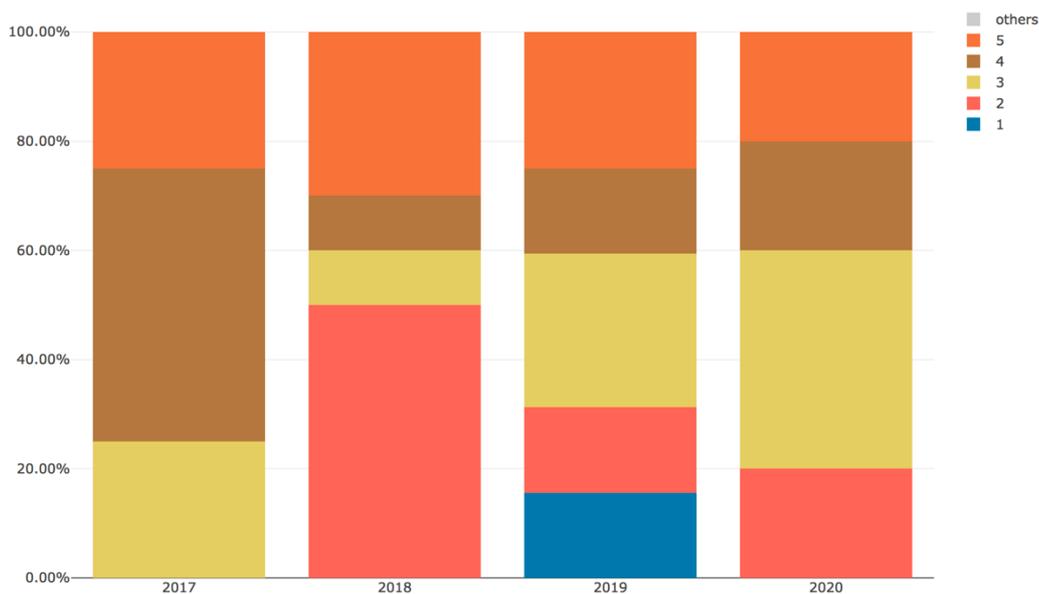

**Figure 3. The MTMvis visualization of the topic model built over a collection of abstracts of articles published in "Umanistica Digitale". We plotted the distribution of the topics according to the year of publication of the articles in the corpus.**

## 4. CONCLUSIONS

In this article, we have claimed the need of supporting scholars in the Humanities having no or limited skills in coding in the use of computational tools for automated textual analysis. In particular, we have presented MITAO, a web-based tool that allows the definition of a visual workflow which can embed several textual analysis activities and methods such as Topic Modelling. MITAO enables the integration of text analysis operations without needing a strong knowledge about their technical implementation and that it builds a visually comprehensive workflow which might be shared with other colleagues and foster the reproducibility of the results obtained. While there is still room for further improvements, the first release of MITAO was already presented and tested during a symposium organized within the EURAM 2019 Conference – Exploring the future of management, which took place in Lisbon (Portugal) in June 2019 (http://pastconferences.euram.academy/programme2019/symposia.html).

In the future, we plan to organise user testing sessions, dedicated workshops, and tutorials on the use of MITAO. This will help us promote our tool in different disciplines, to improve its usability, and to add new relevant features to address particular studies.